\documentclass[twocolumn,showpacs,preprintnumbers,prl,amsmath,amssymb]{revtex4}

\usepackage{graphicx}
\usepackage{dcolumn}
\usepackage{bm}

\begin{document}

\title{Antiferromagnetism in the magnetoelectric effect single crystal LiMnPO$_4$}

\author{Jiying Li$^{1,2}$,  Wei Tian$^3$, Ying Chen$^{1,2}$, Jerel L. Zarestky$^3$, Jeffrey W. Lynn$^1$, and David Vaknin$^3$\footnote{electronic mail: vaknin@ameslab.gov}}

\affiliation{$^1$NIST Center for Neutron Research, National Institute of Standards and Technology, Gaithersburg, MD 20899\\
$^2$Department of Materials Science and Engineering, University of Maryland, College Park, MD 20742\\
$^3$Ames Laboratory and Department of Physics and Astronomy, Iowa State University, Ames, IA 50011}

\date{\today}

\begin{abstract}
Elastic and inelastic neutron scattering studies reveal details of the antiferromagnetic tansition and intriguing spin-dynamics in the magneto-electric effect single crystal LiMnPO$_4$.  The elastic
scattering studies confirm the system is antiferromagnetic (AFM) below $T_N$=33.75 K with local magnetic
moments (Mn$^{2+}$; $S = 5/2$) that are aligned along the crystallographic $a$-axis. The spin-wave dispersion curves propagating along the three
principal axes, determined by inelastic scattering, are adequately modeled in the linear spin-wave framework
assuming a spin-Hamiltonian that is parameterized by inter- and in-plane nearest- and next-nearest-neighbor interactions,
and by easy-plane anisotropy.  The temperature dependence of the spin dynamics makes this an excellent model many-body spin system to address
the question of the relationship between spin-wave excitations and the order parameter.

\end{abstract}

\maketitle

\section{Introduction }
\begin{figure} [htl]
\centering
\includegraphics[width = 0.43\textwidth] {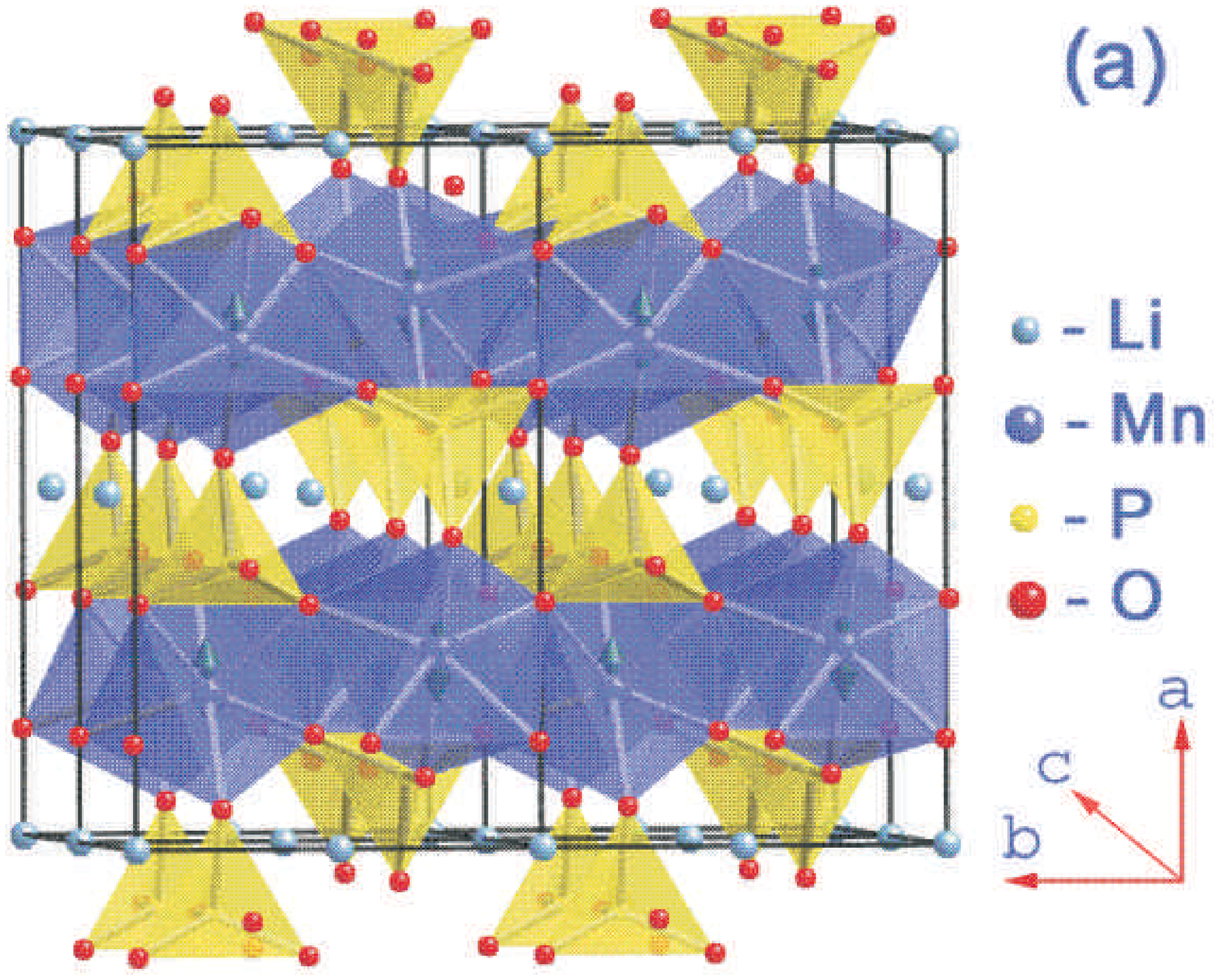}
\includegraphics[width = 0.40\textwidth] {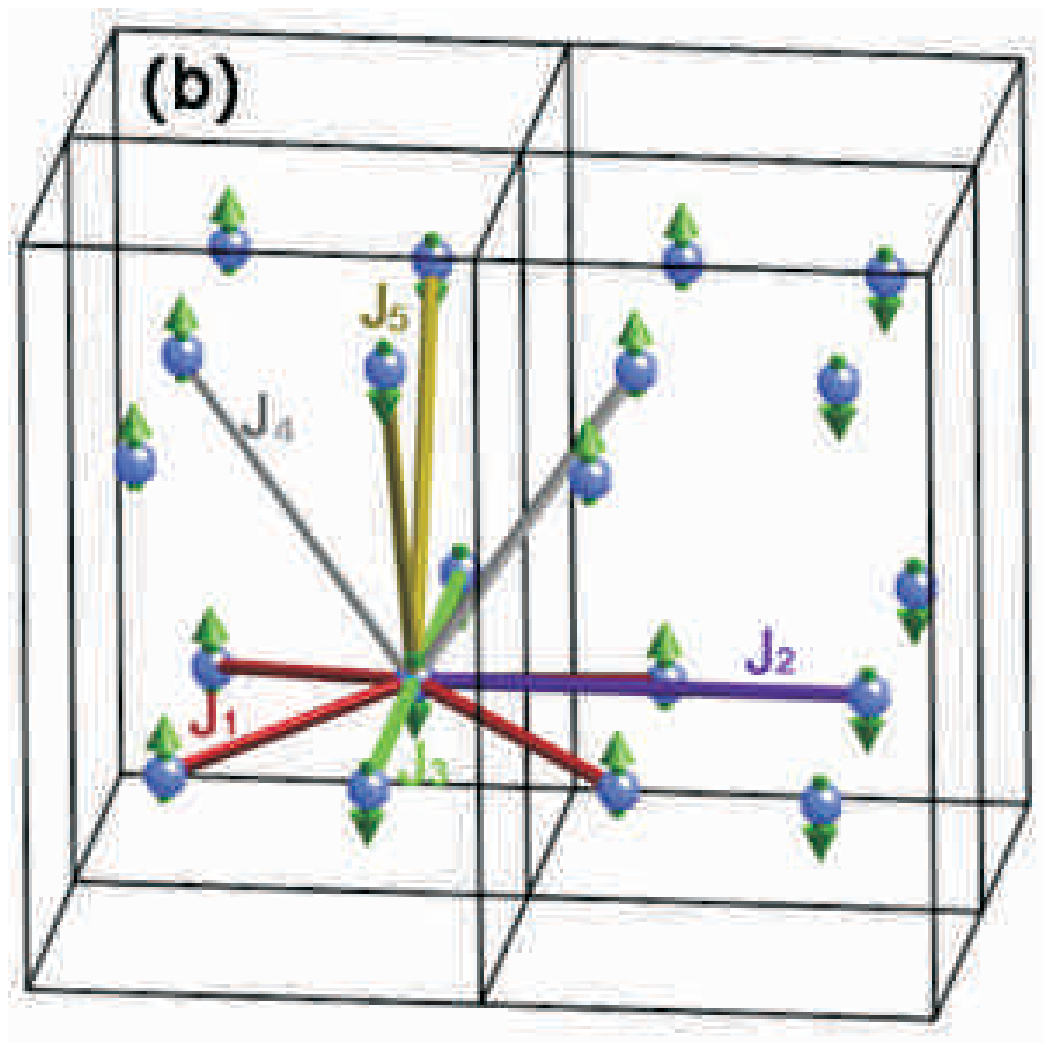}
\caption{\label{Str}(color online) (a) Atomic structure of LiMnPO$_4$. The
Mn$^{2+}$ ions form buckled layers stacked perpendicular to the [100]
crystallographic direction. The ground state of LiMnPO$_4$ is collinear
antiferromagnetic with average moments along the {\it a}-axis. (b) Spin
arrangement of the two Mn$^{2+}$ layers. The in-plane nearest and next-nearest
neighbor interactions {\it J}$_1$, {\it J}$_2$, {\it J}$_3$ and inter-plane
nearest and next-nearest neighbor interactions {\it J}$_4$, {\it J}$_5$ are
labeled.} \label{structure}
\end{figure}

The recent discoveries of colossal magnetoelectric effects (ME) in
rare-earth-manganites (RMnO$_3$ \cite{Kimura2003, Goto2004}) and
manganese-oxides (R$_2$MnO$_5$ \cite{Hur2004}) triggered a revival interest in
the so-called insulating mulitiferroic materials that exhibit ferroelectricity
in coexistence with ferromagnetism or antiferromagnetism (FM or AFM)
\cite{Fiebig2005,Eerenstein2006}.  Systematic studies of the coupling between
the electric and magnetic fields in crystals date back to the early 1960s with
the discovery of the first ME compound Cr$_2$O$_3$ \cite{Astrov1960,Rado1961}.
Early on, the isostructural transition-metal lithium-orthophosphates Li{\it
M}PO$_4$ ({\it M} = Mn, Fe, Co, Ni) were identified as ME systems
\cite{Mericer1967,Mericer1968,Rivera1994} and have been the subjects of
numerous studies \cite{Schmid1973,Schmid1993,Fiebig2004}.  Like other members
of the lithium-orthophosphates, LiMnPO$_4$ is an antiferromagnetic insulator
with {\it Pnma} symmetry group \cite{Geller1960,Megaw1973}.  In this structure,
each Mn$^{2+}$ ion occupies the center of a slightly distorted MnO$_6$
octahedron that shares oxygen anions with a tetrahedral PO$_4$ forming a
closely packed oxygen framework.  The Mn$^{2+}$ ions ($S = 5/2$) form buckled
layers that are stacked along the [100] crystallographic axis, as shown in Fig.
\ref{structure}(a).  The nearest neighbors (NN) in the {\it b-c} plane are
coupled magnetically by a relatively strong exchange interaction {\it J}$_1$
through an Mn-O-Mn oxygen-bond, whereas the in-plane next-NN (NNN) are coupled
via Mn-O-O-Mn ({\it J}$_2$) \cite{Mays1963,Dai2005} (see Fig.
\ref{structure}(b) for the definitions of the exchange couplings). The
interlayer magnetic coupling is mediated through phosphates by higher order
superexchange via Mn-O-P-O-Mn, which was found to be relatively large in
similar frameworks \cite{Zarestky2001}.

Neutron diffraction of polycrystalline samples
\cite{Santoro1966,Santoro1967,Vaknin1999} and single crystal NMR
\cite{Mays1963}  measurements showed that all Li$M$PO$_4$ share the same
collinear (up-down) AFM ground state with spin orientation along {\it a}, {\it
b}, {\it b} and {\it c} crystallographic directions for LiMnPO$_4$, LiFePO$_4$,
LiCoPO$_4$ and LiNiPO$_4$, respectively.  However, recent single crystal
neutron diffraction studies of LiCoPO$_4$, LiFePO$_4$, and LiNiPO$_4$,
\cite{Vaknin2002,Li2006,Tian2008,Jensen2009a} show the moments in the ground state
are slightly tilted away from principal crystallographic directions, indicating the magnetic symmetries for these systems are lower
than those determined from polycrystalline measurements, giving rise to
spontaneously induced weak ferromagnetism.  Weak ferromagnetism (WFM) in
magnetic susceptibility measurements has also been reported for LiNiPO$_4$
\cite{Kharchenko2003} and LiMnPO$_4$ \cite{Arcon2004} below {\it T}$_N$.
Indeed, domain structures observed by second-harmonic-generation (SHG)
experiments in LiCoPO$_4$ were interpreted as ferrotoroidic domains
\cite{VanAken2007} facilitated by the lower magnetic symmetry obtained in
neutron scattering experiments \cite{Vaknin2002}.  Based on the detailed spin configuration
observed in LiNiPO$_4$, Jensen and co-workers
have been able to model the temperature dependence of the ME coefficients of
this system \cite{Jensen2009a}.

Here, we report elastic and inelastic neutron scattering studies of a single
crystal LiMnPO$_4$, to determine the
nature of the AFM transition and the spin dynamics in this system.  Recent
susceptibility measurements indicated WFM in this system \cite{Arcon2004}
implying spin-canting that may be detected in neutron diffraction measurements.
There is also some inconsistency in the literature with regard to the
transition temperature; $T_N = $34.85 \cite{Santoro1966,Santoro1967} and  42 K
were reported \cite{Arcon2004} for polycrystalline samples.   The spin dynamics
of the LiFePO$_4$, LiCoPO$_4$ and LiNiPO$_4$ were measured and modeled in the
linear spin-wave framework only recently \cite{Li2006,Tian2008,Jensen2009b}, to
successfully yield the exchange couplings and the single-ion anisotropy parameters in
these systems. Determining and analyzing the spin dynamics of LiMnPO$_4$ is an
important step towards developing a universal understanding of the magnetic
properties of this isostructural group of compounds.

\section{Experimental Details}
A LiMnPO$_4$ single crystal (0.41 gram, pink in color) was grown by the
standard flux growth technique (LiCl was used as the flux) from a
stoichiometric mixture of high purity MnCl$_2$ (99.999$\%$ Aldrich) and
Li$_3$PO$_4$ (99.999$\%$ Aldrich) \cite {Fomin2002}. Powder, for the XRD, was
prepared by crushing typical isolated single crystals. The composition and
structure were confirmed by carrying out Rietveld analysis of X-ray powder
diffraction (XRD) data, using the GSAS software package \cite{Larson1990}.  No
extra peaks from impurities were detected in the XRD pattern.  The lattice
parameters yielded from the refinement at room temperature ($a = 10.524 $
{\AA}, $b = 6.095 $ {\AA}, and $c = 4.75 $ {\AA}) are in good agreement with
the values reported in the literatures \cite
{Santoro1967,Streltsov1993,Rousse2003}.

Neutron scattering measurements were carried out on the BT7 and BT9 thermal
triple axis spectrometer at the National Institute of Standards and Technology (NIST)
Center for Neutron Research (NCNR). A monochromatic neutron beam of wavelength
$\lambda $ = 2.36 \AA\ (14.7 meV, $k_{o}=2\pi /\lambda=2.66$\AA$^{-1}$) was
selected by a vertical focusing monochromator system, using the (0 0 2) Bragg
reflection of highly oriented pyrolytic graphite (HOPG) crystals. HOPG crystals
were also used as analyzer for both the elastic and the inelastic studies. The
high resolution inelastic scattering measurements were conducted on the on the
cold neutron Spin Polarized Inelastic Neutron Spectrometer (SPINS) at the NCNR.

\section{Results and Discussion \protect}
\subsection{Elastic neutron scattering}
The LiMnPO$_4$ crystal was oriented with its {\it a-b} plane, (and subsequently
rotated with its {\it b-c} plane) to coincide with the scattering-plane of the
spectrometer.  The elastic measurements, with the strongest magnetic reflection
(010) peak, confirmed that the magnetic structure of LiMnPO$_4$ is AFM with
spin orientation along the {\it a}-axis.
\begin{figure}[ht!]
\centering
\includegraphics[width = 0.43\textwidth] {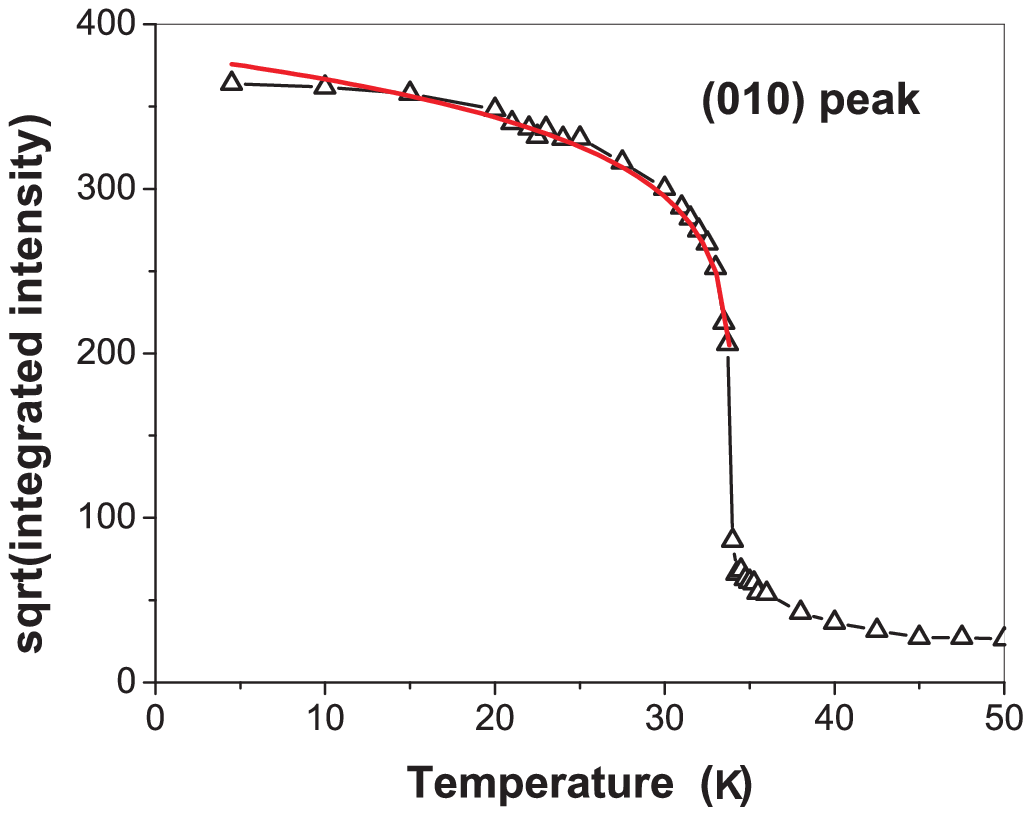}
\caption{(Color online) Temperature dependence of the square root of the
integrated intensity of the (010) magnetic peak. The transition temperature
obtained from the fit is {\it T}$_N$ = 33.85 $\pm$ 0.1 K and the critical
exponent $\beta$ = 0.126 $\pm$ 0.017.} \label{Order}
\end{figure}

The temperature dependence of the magnetic (010) reflection was
used to extract the behavior of the order parameter in the temperature range
from 5 to 50 K.  Figure \ref{Order} shows the square root of the integrated
intensity, representing the staggered magnetization, (i.e., order parameter) as
a function of temperature for the (010) peak.  The order parameter
was fit to a power law function near the transition temperature:
\begin{equation}
\sqrt{I} \mbox{ }{\propto}\mbox{ } M^{\dagger} =
M^{\dagger}_0t^{\beta} \label{eq3}
\end{equation}
where {\it M}$^\dag_0$ is the sublattice magnetization at {\it T} = 0 K, {\it
t} = (1-{\it T}/{\it T}$_N$) is the reduced temperature, and $\beta$ is the
critical exponent.  The obtained transition temperatures from the fit is {\it
T}$_N$ = 33.85 $\pm$ 0.1 K and the critical exponent for the temperature
dependent magnetization $\beta$ is 0.126 $\pm$ 0.017 by using the main (010)
magnetic peak.  This is very close to the theoretical value of the critical exponent of the  2D Ising system $\beta$ =0.125 \cite{Collins1989}, consistent with the layered nature of the magnetic system, as also demonstrated by the weak interlayer coupling obtained from the analysis of the spin-waves discussed below.
The transition temperature is found to be  very close to the value, 34.85 $\pm$ 0.1 K measured by Mays \cite{Mays1963}
using nuclear magnetic resonances performed on a single crystal of LiMnPO$_4$,
whereas susceptibility measurements of powder samples  yield $T_N$  = 42 K
\cite{Arcon2004}.

\begin{figure}[ht!]
\centering
\includegraphics[width = 0.43\textwidth] {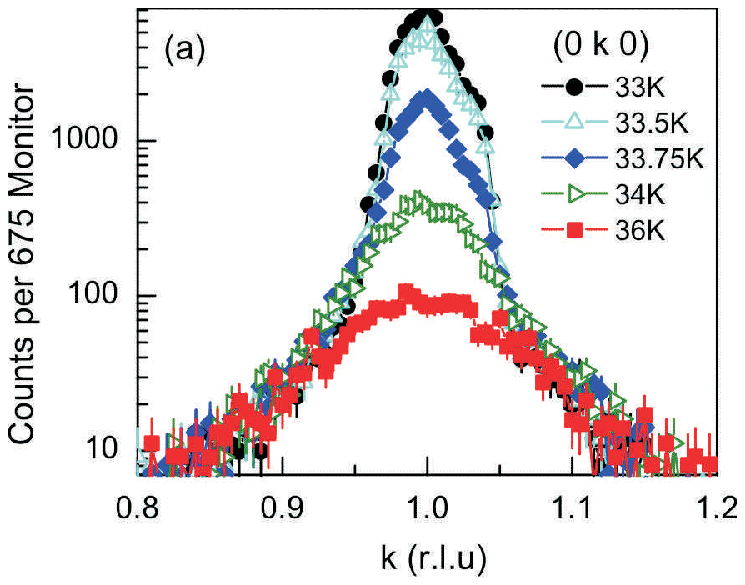}
\includegraphics[width = 0.43\textwidth] {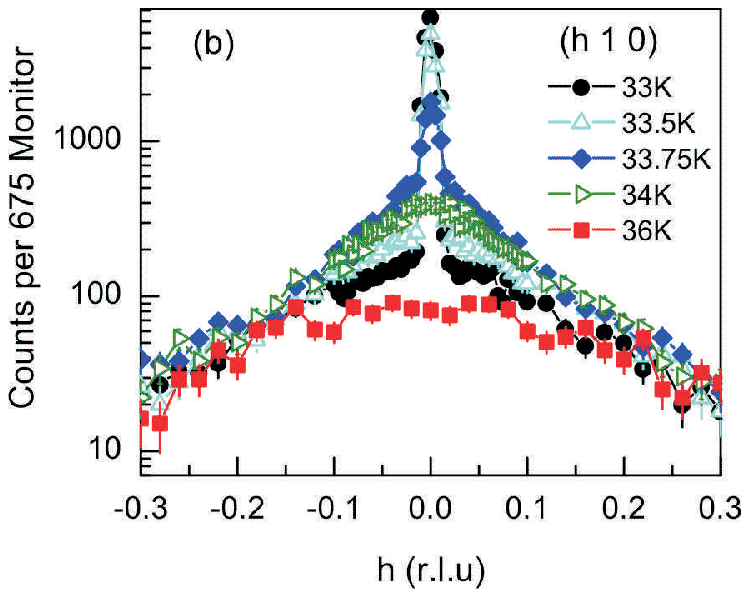}
\includegraphics[width = 0.43\textwidth] {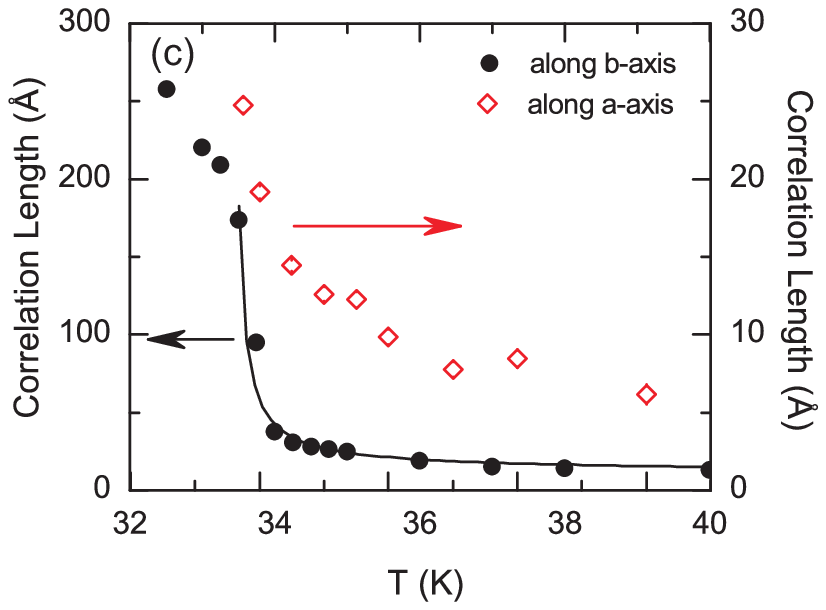}
\caption{(Color online) (a) Longitudinal and (b) transversal, (i.e.,
inter-plane) critical magnetic scattering scans  at the (010) magnetic reflection
above $T_N$.  (c) Correlation lengths obtained after deconvoluting the
spectrometer's resolution function (in-plane and inter-plane as indicated)
versus temperature.  Solid line is calculated assuming Kosterlitz-Thouless
theory. The error bars in this paper are statistical in origin and represent one
standard deviation. (r.l.u. stands for reciprocal lattice unit)} \label{coherence}
\end{figure}

Unlike LiCoPO$_4$ and LiFePO$_4$, strong critical scattering above the AFM
transition is observed in LiMnPO$_4$ and persists to almost twice $T_N$ (traced
to temperatures as high as 70 K) before the spins become uncorrelated. These
correlations were already evident in Figure \ref{Order}.  Figure
\ref{coherence}  shows longitudinal and transversal scans at the (010)
magnetic peak above the transition with energy transfer $\bigtriangleup E$
= 0.  The peaks are much broader than the spectrometer's resolution indicating
some type of short range correlations.  This is reminiscent of the behavior in
LiNiPO$_4$ where this critical scattering \cite{Vaknin1999} was later found to
be associated with an incommensurate (IC) short- and long-range magnetic order
above $T_N$ \cite{Vaknin2004}. Below $\sim$33 K all peaks are practically
resolution limited Gaussian shaped. Above $T_N$, the peaks  were
fit to a Lorentzian line shape $1/(q^2 + \kappa^2)$, where $q = h$ or $k$, and
$\kappa$ is inversely proportional to the coherence length $\xi =2\pi/\kappa$,
convoluted with an instrumental gaussian shaped resolution function. The calculated coherence
lengths along the $a$- and $b$-axis, as a function of temperature are shown in
Figure \ref{coherence} (c).  Below the transition temperature, the in-plane
coherence lengths (along the b-axis) is significantly longer than that between
the planes (along the $a$-axis), consistent with the quasi-2D nature of
LiMnPO$_4$. The correlation lengths were remeasured on BT9 using the integrated energy (two-axis mode) method and yielded similar results.  Attempts to correlate the in-plane coherence length in the critical regime with the 2D-Ising model failed, but the Kosterlitz-Thouless (KT) 2D XY-model,
$\xi(T) = A e^{B/(T-T_c)^\upsilon}$, seems to fit our data well \cite{Ding1990}
(solid line, Fig.\ \ref{coherence}) with $T_c = 33.6 \pm 0.008$ and
$\upsilon = 0.51 \pm 0.1$. This strong critical scattering above the transition with KT characteristics may therefore indicate spin-dimensionality crossover from the 2D Heisenberg to
the 2D XY-model.

\subsection{Inelastic and quasielastic neutron scattering}

\begin{figure}[ht!]
\centering
\includegraphics[width = 0.43\textwidth] {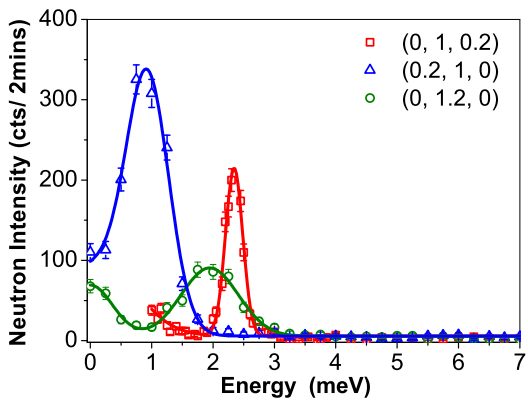}
\caption{(color online) Examples of constant-Q energy scans measured on BT7
at 5 K, at wave-vectors $q$ =0.2 along ($q$, 1, 0), (0, 1+$q$,0) and
(0,1,$q$) reciprocal directions. A single energy excitation is present in
every direction with the typical energy resolution of BT7 around 1 meV.}
\label{EnergyExcite}
\end{figure}

\begin{figure}[htb!]
\centering
\includegraphics[width = 0.43\textwidth] {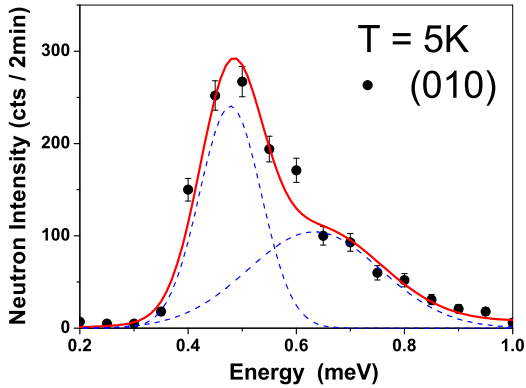}
\caption{(color online) The energy excitation at the zone center measured on
SPINS cold neutron triple axis which has an energy resolution of $\sim$ 0.1
meV. Two excitations are clearly identified at the zone center.}
\label{ExcitatiGap}
\end{figure}

\begin{figure} [htb!]
\centering
\includegraphics[width = 0.43\textwidth] {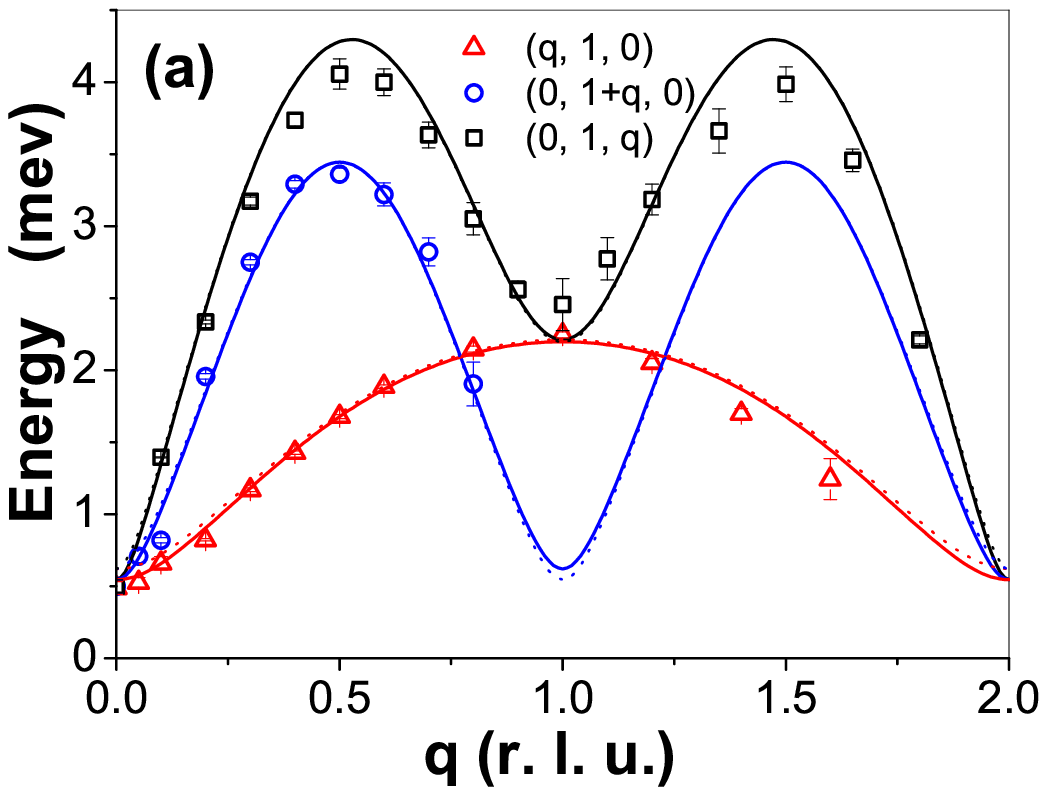}
\includegraphics[width = 0.43\textwidth] {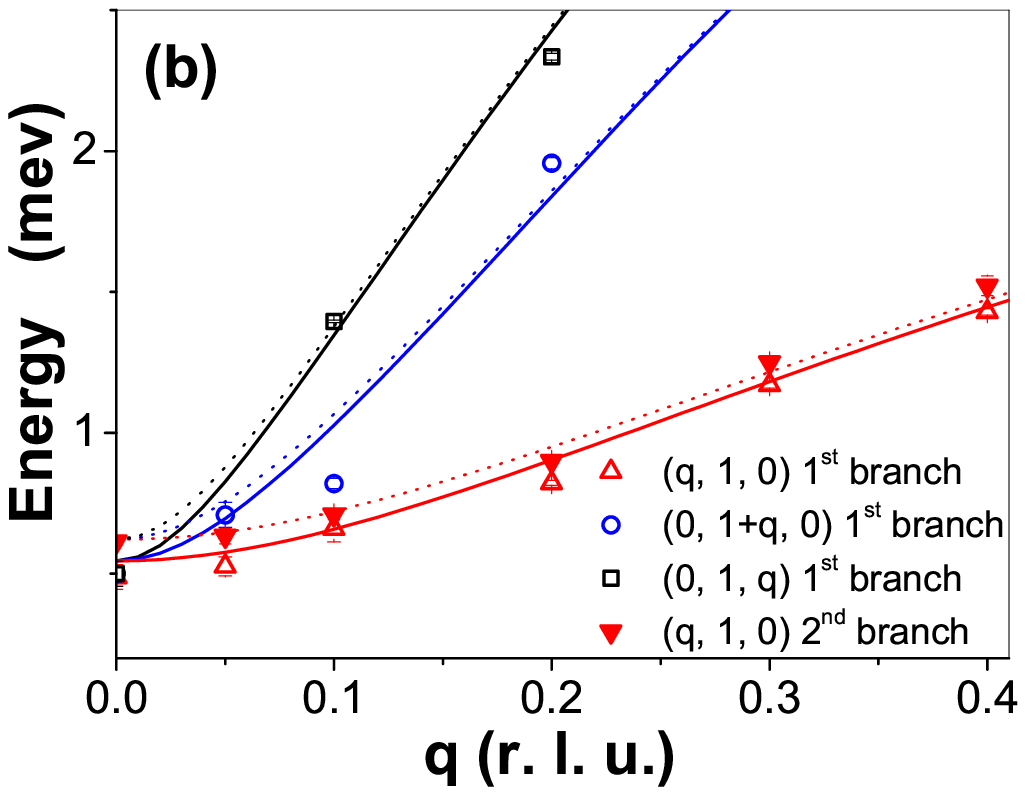}
\caption{\label{dispersion}(color online) (a)Spin-wave dispersion curves along
the {\it a}$^*$, {\it b}$^*$, and {\it c}$^*$ reciprocal space directions
measured at 5 K. Solid lines are best-fit calculations obtained from linear
spin-wave theory using Eq. (\ref{Maineq}). (b) Zoomed plot of (a) near the zone
center. The predicted second spin wave dispersion branches are shown as dashed
lines.}
\end{figure}

Spin waves along the three principal reciprocal lattice directions ($q$,1, 0),
(0, 1+$q$, 0) and (0, 1, $q$) were measured in energy loss mode at T = 5 K.
Examples of the excitations measured on BT7 at $q$ = 0.2 are shown in Figure
\ref{EnergyExcite}.  A single excitation was observed at each $q$ along the three
directions on BT7 which has an energy resolution of $\sim$1 meV. At $T
\approx$ 50 K, no similar peaks are observed confirming the magnetic origin of
the excitations.   The inelastic signals at various constant wave-vectors $q$
were fit to Gaussian shaped functions (solid line in Fig. \ref{EnergyExcite}),
and the set of energies at maximum intensity were used to construct the
spin-wave dispersion curves shown in Fig.\ \ref{dispersion}.  It is shown that
the spin-waves propagating in the plane along the (001) and (010) directions
have higher energy than the spin-waves propagating along (100) at the same $q$
values.  Qualitatively, this behavior reflects the anisotropy in the strength
of the exchange couplings in the system;  as expected, the in-plane exchange
couplings are much stronger than those between planes.  Using the cold neutron
triple axis SPINS spectrometer, an energy gap $E_G = 0.48$ meV was observed around the (010) zone center,  which is much smaller than the 2 meV \cite{Jensen2009b}, 5.86 meV
\cite{Li2006}, and 4.7meV \cite{Tian2008} observed in LiNiPO$_4$, LiFePO$_4$ and
LiCoPO$_4$, respectively.  With the high energy resolution of SPINS,
which is around 0.1 meV (using 3.7 meV final energy), two energy excitation peaks
were identified at the zone center, as shown in Figure \ref{ExcitatiGap}.

To model the spin-wave dispersions, we use a spin Hamiltonian based on the
ground state spin structure of LiMnPO$_4$ as shown in Fig.\ \ref{structure}, as follows
\begin{equation}
{\cal H} = \sum_{i,j}(J_{\{i,j\}}\textbf{S}_i\cdot\textbf{S}_j)+ \sum_{i,\xi}D_{\xi}(S^{\xi}_i)^2
\label{Hamiltonian}
\end{equation}
$J_1$ to $J_5$ are the spin coupling constants (see Fig.\ \ref{structure}), and
D$_\xi$ are the single ion anisotropies.  Since the excitation spectrum is insensitive to an overall shift of the ground state energy we can define  $D_z\equiv 0$ for simplicity.  The $x, y$, and $z$ coordinates are
defined along the $c$-, $b$- and $a$-axis, respectively, to align the spin
direction in the ground state with the quantum $z$-axis in Eq.\
(\ref{Hamiltonian}).  The magnon dispersion curves derived from Eq.
(\ref{Hamiltonian}) by linear spin-wave theory is given in Refs
\cite{Squires1978,Li2006,Jensen2009b}. In the model, the calculated spin waves
have two non-degenerate branches (denoted by the $\pm$ sign in Eq. \ref{Maineq}
) as a result of the different anisotropies along the x and y directions.

\begin{equation}
\hbar\omega = \sqrt{A^2-(B \pm C)^2}
\label{Maineq}
\end{equation}

where,
\begin{eqnarray}
 A\equiv &&4S(J_1+J_5)-2S[J_3(1-\cos(\textbf{q}\cdot {\textbf{r}}_5))+J_2(1- \nonumber\\
 &&\cos(\textbf{q}\cdot {\textbf{r}}_6))+J_4(2-\cos(\textbf{q}\cdot {\textbf{r}}_7)-\cos(\textbf{q}\cdot {\textbf{r}}_8))] \nonumber\\
 &&+(S-1/2)(D_x + D_y), \nonumber
\end{eqnarray}
\begin{equation}
B \equiv (S-1/2)(D_x - D_y), \nonumber
\end{equation}
\begin{eqnarray}
C \equiv &&2J_1S[\cos(\textbf{q}\cdot {\textbf{r}}_1)+\cos(\textbf{q}\cdot {\textbf{r}}_2)]\nonumber\\
&&+2J_5S[\cos(\textbf{q}\cdot {\textbf{r}}_3)+ \cos(\textbf{q}\cdot {\textbf{r}}_4)], \nonumber
\end{eqnarray}
and $\textbf{r}_i$ denotes a vector to a NN and NNN,
\begin{eqnarray}
\textbf{r}_1 = (c/2, b/2, 0) & \textbf{r}_2 = (-c/2, b/2, 0)  \nonumber \\
\textbf{r}_3 = (0, b/2, a/2) & \textbf{r}_4 = (0, -b/2, a/2)  \nonumber \\
\textbf{r}_5 = (0, b, 0)     & \textbf{r}_6 = (c, 0, 0)       \nonumber \\
\textbf{r}_7 = (c/2, 0, a/2) & \textbf{r}_8 = (-c/2, 0, a/2). \nonumber
\end{eqnarray}

The spin-wave dispersion curves along the three directions in Fig.
\ref{dispersion} were simultaneously fit to Eq. (\ref{Maineq}), using the ``$-$''
sign, yielding the following values: {\it J}$_1$ = 0.48 $\pm$ 0.05 meV, {\it
J}$_2$ = 0.2 $\pm$ 0.038 meV, {\it J}$_3$ = 0.076 $\pm$ 0.004 meV, {\it J}$_4$
= 0.036 $\pm$ 0.002 meV, {\it J}$_5$ = 0.062$\pm$ 0.003 meV, {\it D}$_x$ =
0.0069 $\pm$ 0.001 meV and {\it D}$_y$ = 0.0089 $\pm$ 0.001 meV.  In the
equation, $S = 5/2$ for Mn$^{2+}$. As expected, the in-plane NN exchange
coupling $J_1$, is the strongest, compared to the in-plane NNNs $J_2$ and
$J_3$. The sign of both $J_2$ and $J_3$ indicates the NNN interactions compete
with the simple AFM ordering dictated by $J_1$. For weakly coupled layers, it has been predicted theoretically that an incommensurate (IC) magnetic structure should be realized when $J_2/J_1 > 0.5$\cite{Nagamiya1967}. Thus, unlike in LiNiPO$_4$  where the $J_2/J_1 \approx 0.6$ and an IC has been observed, the ratio for LiMnPO$_4$ ($\sim$ 0.4) seems to be too small to induce any IC phase transition \cite{Jensen2009b}.  The spin
couplings between the inter-plane nearest-neighbors ($J_4$ and $J_5$) are relatively
weak at about 12\% of $J_1$ consistent with the quasi-2D behavior of
this system.  The values of the single ion $D_x$ = 0.0055  and $D_y = 0.0071$,
are much smaller than those of LiNiPO$_4$ \cite{Jensen2009b}, LiFePO$_4$
\cite{Li2006}, and LiCoPO$_4$ \cite{Tian2008} indicating that the ground state
with magnetic moments along the $a$-axis is not very stable, and the moments are prone to a spin-flop transition in relatively weak magnetic fields \cite{Ranicar1967,Elliston1969}.

The second spin wave dispersion branches, given by ``+'' sign in Eq. (\ref{Maineq}), are calculated using the $J$s and
$D$s obtained from the fits listed above. The two branches almost overlap one
another for the dispersions along all the three principal reciprocal
directions, and are only separate by $\sim$ 0.1 meV at the zone center. The
spin wave dispersion along ($q$, 1, 0) direction, where the model predicts the
largest separation between the two branches, was remeasured with the high
energy resolution on SPINS. Figure \ref{dispersion} (b) shows an enlargement graph of
Figure \ref{dispersion} (a), with fairly good agreement with the model calculations.

\begin{figure}[hbt!]
\centering
\includegraphics[width = 0.43\textwidth] {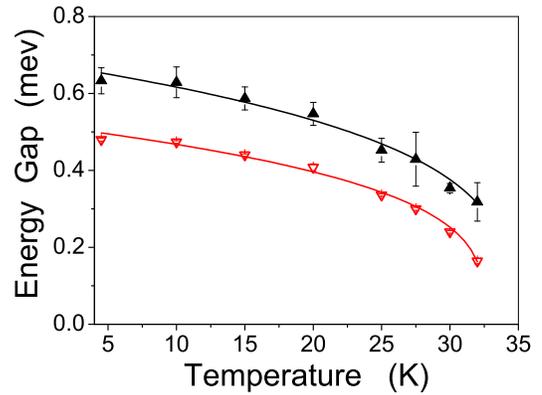}
\caption{(Color online)  The energy gap as a function of temperature measured
on the SPINS cold neutron triple axis.}
\label{energygap}
\end{figure}

The energy gaps at the zone center for the two branches are
\begin{equation}
\Delta E = 2S\sqrt{4D_x(J_1+J_5)+D_xD_y},
\label{EngergyGap}
\end{equation}
for (B - C) in Eq. (\ref{Maineq}) and
\begin{equation}
\Delta E = 2S\sqrt{4D_y(J_1+J_5)+D_xD_y},
\label{EngergyGap2}
\end{equation}
for (B + C).  {\it J}$_5$ represents the inter-plane NN coupling.  From the
equations, we notice that the energy gap not only depends on the single-ion
anisotropy terms, but also on the two nearest-neighbor antiparallel exchange
interactions.

The temperature dependent energy gap up to the transition temperature was
measured at the cold neutron triple axis spectrometer SPINS, and the results
are shown in Figure \ref{energygap} (energy gaps at various temperatures were determined from gaussian fits to constant-Q energy scans such as the one shown in Fig.\ \ref{ExcitatiGap} at $T=5$ K).  The energy gap monotonically decreases
with increasing temperature and approaches zero at the transition temperature.
The temperature dependence of the gap to a first approximation is proportional
to the staggered magnetization which is temperature dependent\cite{Bloch1962}. However, it
may deviate in the critical regime due to the different temperature
dependencies of the coupling constants and the single ion anisotropy.  In
antiferromagnets, the exchange constants $J_s$ usually decrease much faster
than the single ion anisotropy near the transition temperature \cite{Nagai1969,
Bloch1962}.

\begin{figure}[htb!]
\centering
\includegraphics[width = 0.43\textwidth] {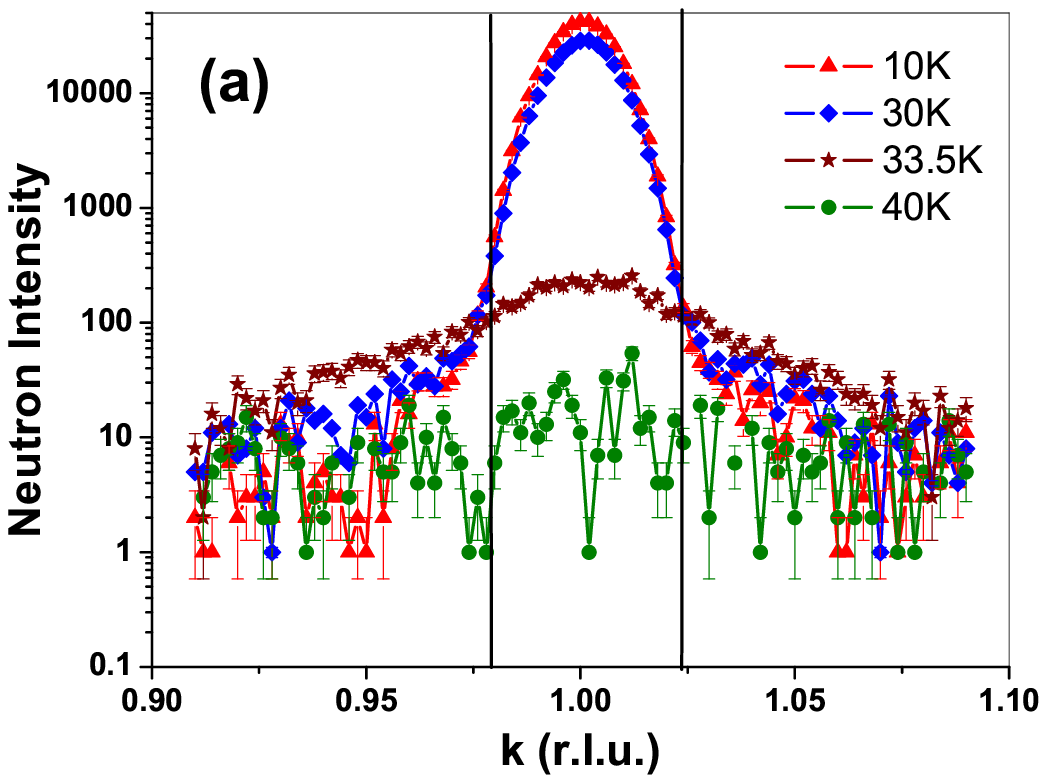}
\includegraphics[width = 0.43\textwidth] {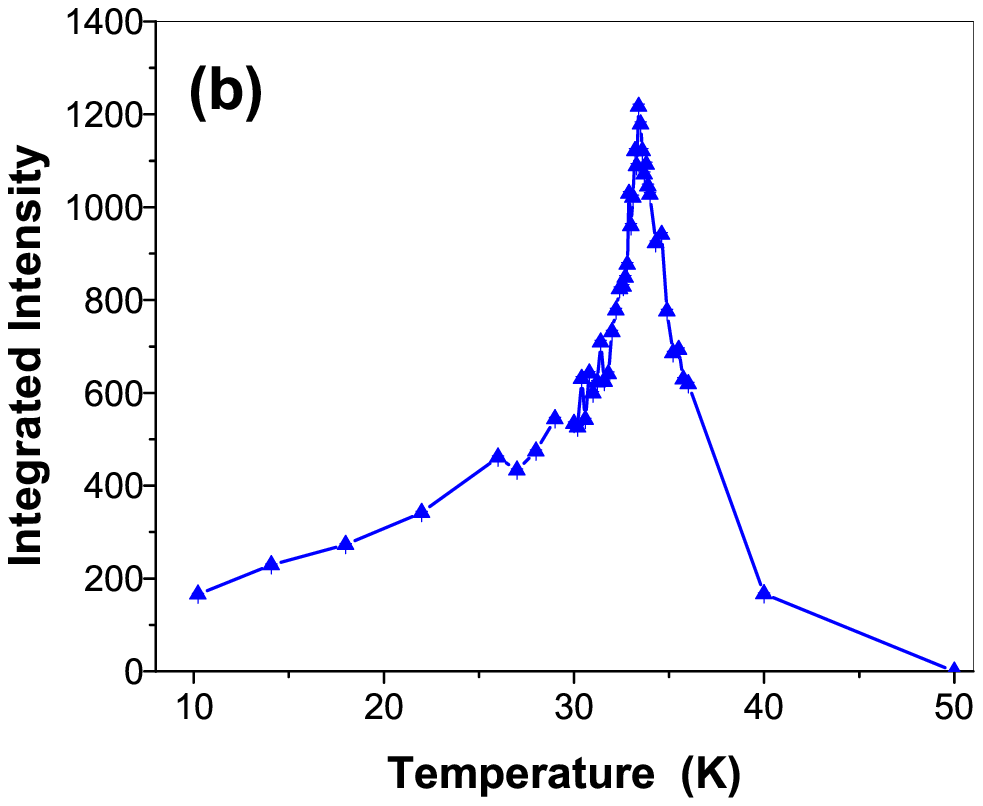}
\caption{(Color online)(a) Examples of the quasielastic scattering scans around
(010) magnetic peak at temperatures below and above the transition temperature
$T_N$, which were measured on BT9 using the integrated energy method. (b) The
temperature dependent integrated intensity from the quasielastic scattering
excluding Q = 0.98 to 1.02 as indicated by the box in (a).}
\label{Quasielastic}
\end{figure}

Quasi-elastic scattering (QENS) around (010) at different temperatures was measured on BT9 using the integrated energy (two-axis) mode, and the results are shown in
Figure \ref{Quasielastic}. At temperatures right below the transition, the
(010) peak consists of a resolution limited Gaussian shaped magnetic Bragg peak
superimposed on a broad Lorentzian shaped diffuse peak.  Whereas the diffuse scattering
becomes stronger with the increase of temperature (up to the transition), the elastic magnetic Bragg peak becomes weaker.  The QENS intensity at each temperature was integrated over the
the K range shown in Figure\ \ref{Quasielastic} (a) excluding the region from 0.98 to 1.02 (r.l.u) which is dominated by elastic scattering.   Figure \ref{Quasielastic} (b) shows the QENS as a function of temperature, which exhibits a sharp peak at the transition ($T_N$ =
33.75 K) with a tail that extends to about 1.5$T_N$.  This indicates that the short
range correlations observed in the elastic scattering are primarily due to (dynamics)
spin-fluctuations.

In summary, we determined the critical behavior near the AFM magnetic phase transition of LiMnPO$_4$ ($T_N$ = 33.85 K). The strong critical scattering around the (010) magnetic peak and the in-plane, inter-plane coherence lengths indicate that the system is a quasi-2D system
with very weak easy axis single ion anisotropy.  Analysis of the spin-wave
dispersions along the three principal axis directions show that the in-plane
couplings are dominant compared to the inter-plane couplings. These in-plane
competing interactions between in-plane NN and NNN-spins in LiMnPO$_4$ seem to
be too weak to lead to more complicated, incommensurate magnetic structures.
This is in contrast to the observation of incommensurate magnetic phases in
LiNiPO$_4$ \cite{Vaknin2004}.

\begin{acknowledgments}
We thank Sung Chang for the support on the SPINS cold neutron triple axis. Ames Laboratory is supported by the U.S. Department of Energy, Basic Energy Sciences, Office of Science, under Contract No. DE-AC02-07CH11358. SPINS is supported in
part by the US National Science Foundation through DMR-0454672.

\end{acknowledgments}

\begin{references}
\bibitem{Kimura2003} T. Kimura, T. Goto, H. Shintani, K. Ishizaka, T. Arima, and Y. Tokura, Nature {\bf 426}, 55 (2003).
\bibitem{Goto2004} T. Goto, T. Kimura, G. Lawes, A. P. Ramirez, and Y. Tokura, Phys. Rev. Lett. {\bf 92}, 257201 (2004).
\bibitem{Hur2004} N. Hur, S. Park, P. A. Sharma, S. Guha, and S. W. Cheong, Phys. Rev. Lett. {\bf 93}, 107207-1 (2004).
\bibitem{Fiebig2005} M. Fiebig, J. Phys. D: Appl. Phys., {\bf 38}, R123 (2005).
\bibitem{Eerenstein2006} W. Eerenstein, N. D. Mathur, and J. F. Scott, Nature {\bf 442}, 759 (2006).
\bibitem{Astrov1960} D. N. Astrov, J. Exp. Theoret. Phys. (U.S.S.R.) {\bf 38}, 984 (1960).
\bibitem{Rado1961} G. T. Rado and V. J. Folen, Phys. Rev. lett. {\bf 7}, 310 (1961).
\bibitem{Mericer1967} M. Mercier, and J. Gareyte, Sol. State Comm. {\bf 5}, 139 (1967); {\bf 7}, 149 (1969).
\bibitem{Mericer1968} M. Mercier, and P. Bauer, C. R. Acad. Sci. Paris {\bf267}, 465 (1968); and M. Mercier, Ph.D. thesis, Université de Grenoble, 1969.
\bibitem{Rivera1994} J. -P. Rivera, Ferroelectrics {\bf 161}, 147 (1994).
\bibitem{Schmid1973} {\it Magnetoelectric interaction phenomena in crystals}, edited by A.J. Freeman and H. Schmid (Gordon and Breach Science Publishers, London/New York, 1975); references therein.
\bibitem{Schmid1993} Proceedings of the Second International Conference on Magnetoelectric Interaction Phenomena in Crystals (MEIPIC-2), Ascona, 1993, Parts I and II, edited by H. Schmid, A. Janner, H. Grimmer, J.-P. Rivera, and Z.-G.Ye [Ferroelectrics {\bf 161}, 1 (1994); {\bf 162}, 1 (1994)], references therein.
\bibitem{Fiebig2004} {\it Magnetoelectric Interactions Phenomena in Crystals}, edited by M. Fiebig, V. V. Eremenko, and I. E. Chupis, NATO Science Series (Kluwer Academic Publishers, Dordecht, 2004); references therein.
\bibitem{Geller1960} S. Geller and J. L. Easson, Acta Crystallogr., {\bf 18}, 258 (1960).
\bibitem{Megaw1973} H. D. Megaw, {\it Crystal Structures - A Working Approach}, Saunders, Philadephia, 1973, P249.
\bibitem{Mays1963} J. M. Mays, Phys. Rev. {\bf 131}, 38 (1963).
\bibitem{Dai2005} D. Dai, M. H. Whangbo, H. J. Koo, X. Rocquefelte, S. Jobic, A. Villesuzanne, Inorganic Chem., {\bf 44}, 2407 (2005).
\bibitem{Zarestky2001} J. L. Zarestky, D. Vaknin, B. C. Chakoumakos, T. Rojo, A. Goñi, and G. E. Barberis, J. Mag. Mag. Matt. {\bf {234}}, 401 (2001).
\bibitem{Santoro1966}  R. P. Santoro, R. E. Newnham, and S. Nomura, J. Phys. Chem. Solids {\bf 27}, 655 (1966); R. P. Santoro, D. J. Segal, R. E. Newnham, J. Phys. Chem. Solids {\bf 27}, 1192 (1966).
\bibitem{Santoro1967} R. P. Santoro and R. E. Newnham, Acta. Cryst. {\bf 22}, 344 (1967).
\bibitem{Vaknin1999} D. Vaknin, J. L. Zarestky, J. E. Ostenson, B. C. Chakoumakos, A. Go\~{n}i, P. J. Pagliuso, T. Rojo, and G. E. Barberis, Phys. Rev. B, {\bf 60}, 1100 (1999).
\bibitem{Vaknin2002} D. Vaknin, J. L. Zarestky, L. L. Miller, J. -P. Rivera, and H. Schmid, Phys. Rev. B, {\bf 65}, 224414 (2002).
\bibitem{Li2006} J. Li, V. O Garlea, J. L. Zarestky, and D. Vaknin, Phys. Rev. B {\bf 73}, 024410 (2006).
\bibitem{Tian2008} W. Tian, J. Li, J. W. Lynn, J. L. Zarestky, and D. Vaknin, Phys. Rev. B {\bf 78}, 184429 (2008).
\bibitem{Jensen2009a} T. B. S. Jensen, N. B. Christensen, M. Kenzelmann, H. M. R{\o}nnow, C. Niedermayer, N. H. Andersen, K. Lefmann, J. Schefer, M. v. Zimmermann, J. Li, J. L. Zarestsky, and D. Vaknin, Phys. Rev. B {\bf 79}, 092412 (2009).
\bibitem{Kharchenko2003} Y. Kharchenko, N. Kharchenko, M. Baran and R. Szymczak, Low Temp. Phys. {\bf 29}, 579 (2003).
\bibitem{Arcon2004} D. Ar\v{c}on, A. Zorko, P. Cevc, R. Dominiko, M. Bele, J. Jamnik, Z. Jaglicic, I. Golosocsky, J. Phys. Chem. Solids, {\bf 65}, 1773 (2004); D. Ar\v{c}on, A. Zorko, R. Dominko, Z. Jaglicic, J. Phys. Chem. Solids., {\bf 16}, 5531 (2004).
\bibitem{VanAken2007}  B. van Aken, J.-P. Rivera, H. Schmid, and M. Fiebig, Nature {\bf 449}, 702 (2007).
\bibitem{Jensen2009b} T. B. S. Jensen, N. B. Christensen,M. Kenzelmann, H. M. R{\o}nnow,C. Niedermayer,N. H. Andersen, K. Lefmann,M. Jim{\'e}nez-Ruiz, F. Demmel, J. Li, J. L. Zarestky, and D. Vaknin, Phys. Rev. B {bf 79}, 092413 (2009).
\bibitem{Fomin2002} V. I. Fomin, V. P. Genezdilov, V. S. Kurnosov, A. V. Peschanskii, A. V. Yeremenko, H. Schmid, J. -P. Rivera, and S. Gentil, Low Temp. Phys., {\bf 28}, 203 (2002).
\bibitem{Larson1990} A. C. Larson, R. B. Von Dreele, and M. Lujan Jr., {\it Computer code GSAS, Generalized Structure Analysis System}, Neutron Scattering Center, Los Alamos National Laboratory, 1990.
\bibitem{Streltsov1993} V. A. Streltsov, E. L. Belokoneva, V. G. Tsirelson and N. K. Hansen, Acta Crystal., {\bf B49}, 147 (1993).
\bibitem{Rousse2003} G. Rousse, J R. Carvajal, S. Patoux and C. Masquelier, Chem. Mater., {\bf 15}, 4082 (2003).
\bibitem{Collins1989} M. F. Collins, {\it Magnetic Critical Scattering}, New York, Oxford University Press, 1989, p29.
\bibitem{Vaknin2004} D. Vaknin, J. L. Zarestky, J. -P. Rivera, and H. Schmid, Phys. Rev. Lett., {\bf 92}, 207201/1 (2004).
\bibitem{Ding1990} H. Q. Ding and M. S. Makivi$\acute{c}$, Phys. Rev. B, {\bf 42}, 6827 (1990).
\bibitem{Squires1978} G. L. Squires, in {\it Introduction to the Theory of Thermal Neutron Scattering }, Cambridge University Press, New York, 1978, P156.
\bibitem{Nagamiya1967} T. Nagamiya, {\it Solid State Physics} edited F. Seitz and D. Turnbull (Academic, New York, 1967), Vol. 29, p. 346.
\bibitem{Ranicar1967} J. H. Ranicar and P. R. Elliston, Phys. Lett. {\bf 25A}, 720 (1967).
\bibitem{Elliston1969} P. R. Elliston, J. G. Creer, and G. J. Troup, J. Phys. Chem. Solids {\bf 30}, 1335 (1969).
\bibitem{Nagai1969} O. Nagai, Phys. Rev. {\bf 180}, 557 (1969).
\bibitem{Bloch1962} M. Bloch, Phys. Rev. Lett. {\bf 9}, 286 (1962), and J. Appl. Phys. {\bf 34}, 1151 (1963).
\end{references}

\end{document}